To quote this work :

V. Brien, L.P. Kubin and B. Décamps
Low Cycle Fatigue of a Nickel Based Superalloy at High Temperature: Simplified Micromechanical Modelling
Philosophical Magazine A Vol. 81, n°9 pp 2285-2301 (2001) (IF = 1,632)
https://hal.archives-ouvertes.fr/ hal-02882484, doi.10.1080/01418610108217148

Thank you

# Low Cycle Fatigue of a Nickel Based Superalloy at High Temperature :

## Simplified Micromechanical Modelling

By V. BRIEN[1], L.P. KUBIN[2] and B. DECAMPS[3]

1 Laboratoire de Sciences et Génie des Matériaux Métalliques,
UMR CNRS-Ecole des Mines, UMR CNRS 7584,
Parc de Saurupt, 54042 Nancy Cedex, France
2 Laboratoire d'Etudes des Microstructures, UMR CNRS-ONERA,
UMR CNRS 104, BP 72, 92322 Châtillon Cedex, France
3 Laboratoire de Chimie Métallurgique des Terres Rares, UPR CNRS 209,
2-8, rue Henri-Dunant, 94320 Thiais Cedex, France

Corresponding author:

V. Brien,
 LSG2M (CNRS), Ecole des Mines de Nancy, Parc de Saurupt
54042 NANCY Cedex – FRANCE
Tel: +33-3 83 58 40 78; Fax: +33-3 83 57 63 00; E-mail: brien@mines.u-nancy.fr

**ABSTRACT**
This work is focused on the micromechanical modelling of the low cycle fatigue of the nickel based $\gamma/\gamma'$ superalloy *AM1* at high temperature. The nature of the activated slip systems in the different types of channels of the $\gamma$ phase is analysed, taking into account the combined effects of the applied and internal stresses. The latter are split into two contributions, misfit stresses and compatibility stresses between the elastic $\gamma'$ phase and the elasto-plastic $\gamma$ phase, which are estimated within a simplified composite approach. Internal stresses may induce slip activity and/or be relaxed by it, which results in a complex sequence of slip activation events in the different channels under increasing applied stress. The consideration of these effects leads to a prediction of the nature and distribution of the active slip systems within the channels in [001] tension, compression and during low cycle fatigue. The resulting microstructural behaviour and its consequences regarding the anisotropic nature of the coalescence of the $\gamma'$ precipitates are discussed with respect to the available experimental data.

# 1. INTRODUCTION

During monotonic deformation at high temperature, $\gamma/\gamma'$ single crystalline superalloys containing large volume fractions of the ordered $\gamma'$ phase undergo significant microstructural changes. The anisotropic coarsening of the $\gamma'$ precipitates by diffusion leads to the formation of the well known raft structure (Tien and Copley 1971, Carry and Strudel 1977, Pearson, Kear and Lemkey 1981, Caron and Khan 1983, Fredholm 1987, Nabarro 1987), whose spatial arrangement is quite sensitive to the orientation of the loading axis. It has been shown (Véron 1995, see also Véron, Bréchet and Louchet 1996) that the geometry of the raft structure in monotonic tension or compression is conditioned by the difference in plastic behaviour of the horizontal and vertical channels, as induced by local differences in internal stresses. This heterogeneity of plastic flow in superalloys induces an anisotropic relaxation of the misfit stresses. The latter are, in turn, at the origin of the elastic energy gradients, which drive diffusion processes and lead to anisotropic coarsening.

During cyclic deformation, the situation may seem more complicated, as far as coarsening effects are concerned, since the amplitude and sign of the applied load changes periodically. The results of an experimental study of the nickel based superalloy *AM1* during isothermal Low Cycle Fatigue (LCF) at 950ºC (Brien, 1995) are briefly summed up below. Extensive detail on the experimental techniques and the observed microstructures will be published separately (Brien and Décamps, submitted). In repeated tensile fatigue, two significantly different types of dislocation microstructures are formed, depending on the number of cycles and the total strain amplitude. Anisotropic microstructures corresponding to different active slip systems and total dislocation densities in the different $\gamma$ channels are found at low strain amplitudes (cf. fig. 1-a), all the more as the number of cycles is large. In this domain, type *N* directional coarsening, perpendicular to the loading axis, occurs after a certain number of cycles. As the lattice mismatch of *AM1* is $-10^{-3}$ at 950 °C, this is consistent with observations and theoretical expectations for other superalloys with negative lattice mismatch deformed in monotonic tension (cf. Véron, Bréchet and Louchet 1996). In contrast, for larger strain amplitudes, the dislocation microstructures are found to be quite similar in the different $\gamma$ channels (cf. fig. 1-b). All the types of channels plastically deform in the same manner, which leads to homogeneous coarsening for large numbers of cycles.

FIGS. 1 -a, -b, HERE

In alternate fatigue, the available results show that type *N* coarsening occurs in the *AM1* superalloy (cf. fig. 2). In this case, no theoretical prediction exists. Indeed, the type of coarsening induced in monotonic deformation changes with the sign of the applied stress, going from type N in tension to type P, parallel to the loading axis, in compression.

FIG. 2 HERE

The objective of the present paper is to model the nonuniformity of the dislocation microstructure in terms of the various sources of internal stresses present in the $\gamma/\gamma'$ superalloys, with a view of further predicting the coarsening behaviour during LCF. The two sources of internals stresses considered are i) - the stresses induced by the crystallographic misfit between the $\gamma$ and $\gamma'$ phases and ii) - the internal stresses stemming from the condition of compatible deformation between these two phases. The spatial averages of these two contributions are first estimated within a simplified approach in parts 2 and 3. In part 4, the consequences regarding the activation of different slip systems in the different channels are examined. In particular, it is shown that the slip patterns evolve according to well-defined sequences under increasing applied stress. This leads to a modelling of the microstructural behaviour and deformation mechanisms, which is presented and critically discussed.

## 2. MISFIT STRESSES

The stress calculations presented in parts 2 and 3 for the misfit and compatibility stresses are based on the simplifying hypothesis that these internal stresses are uniform in the channels of the $\gamma$ phase. This is clearly not the case in practice since both contributions to the internal stresses originate from the $\gamma/\gamma'$ interfaces and are minimum in the centre of the channels. Further, the state of stress is very complex at the intersections between channels, a point that will not be dealt with in this study. Detailed calculations of the internal stresses, which are recalled below, were performed in monotonic deformation but the extension to the case of cyclic loading has not been attempted. The objective of the present approach is, therefore, to estimate average stress levels in the different types of channels, under different loading conditions, in order to make semi-quantitative predictions on the nature and number of slip systems that are locally activated.

The lattice mismatch $\delta = 2 (a_\gamma - a_{\gamma'})/(a_\gamma + a_{\gamma'})$ is defined as the reduced difference in the lattice parameters of the $\gamma$ and $\gamma'$ phases, $a_\gamma$ and $a_{\gamma'}$, respectively. As the value of $\delta$ depends on the respective thermal dilatations of the two phases, it is also temperature-dependent. Royer, Bastié, Bellet and Strudel (1995) have measured this dependence in the AM1 by high resolution neutron diffraction. At the temperature of interest here (950ºC), the lattice mismatch is negative ($\delta = -10^{-3}$) and bi-axial compression stresses are induced in the matrix channels.

For the purpose of obtaining a simplified estimate of the average misfit stresses, we make use of a result obtained by Saada (1989) who calculated the stress field of a polyhedral inclusion within the frame of linear, isotropic elasticity. For a cubic inclusion whose faces are perpendicular to the three axes of the direct trihedron ($x_1$, $x_2$, $x_3$) the stress tensor $\sigma_m$ in the material containing the inclusion (here the $\gamma$ matrix) is written, e.g. for the face 3 perpendicular to $x_3$ (cf. fig. 3).

$$\sigma_{11}^m = \sigma_{22}^m = K\varepsilon \; ; \quad \sigma_{12}^m = 0 \; , \tag{1}$$

where $\varepsilon = \delta$ is the deformation of the $\gamma$ phase, $K = G(1+\nu)/(1-\nu)$ and where $G$ and $\nu$ are respectively the shear modulus and the Poisson's ratio of the $\gamma$ phase. Similar results hold for the other faces (cf. fig. 3). With $G$ = 45 GPa (Poubanne 1989, Lisiecki 1992) and $\nu \approx 0.4$ (Priester, Khalfallah and Coujou 1997, Müller, Glatzel and Feller-Kniepmeier 1992), eq. 1 yields a value of 157 MPa for the normal stresses at the interfaces.

FIG. 3 HERE

Computations of the misfit stress tensor by finite element methods for various superalloys can be found in the literature (Glatzel and Feller-Kniepmeier 1989, Ganghoffer, Hazotte, Denis and Simon 1991, Pollock and Argon 1992, Benyoucef 1994, Racine and Hazotte 1993, Feng, Biermann and Mughrabi 1996, Ohashi, Hidaka and Imano 1997). Generally, the results thus obtained predict a state of bi-axial compression stress in the $\gamma$ channels with normal stresses in the range of 100-250 MPa, compatible with the present estimate of 157 MPa. It is then customary to calculate the von Mises stress. Although this stress gives a fair idea of the ability of the $\gamma/\gamma'$ structure to deform plastically, it cannot yield any information about the nature of the activated slip systems. The largest component is found to be in the range of 100-150 MPa, even near the interface and the other components are comparatively much smaller. However, it would be very difficult to figure out the forces on the dislocations using the full analytical expression for this stress field.

Taking these results into account, the following simplifying assumption is made. The internal misfit stresses are uniform in the $\gamma$ channels and the value of the compressive components is taken identical to the maximum value at the interfaces, as given by eq. 1 (157 MPa). The main justification for this simplification is that it allows estimating the Peach-Koehler forces on dislocations, as shown in part 4.

## 3. COMPATIBILITY STRESSES

Mughrabi (1981) has developed a simplified composite approach of the internal stresses that develop during loading in bi-phased materials. This model applies to materials containing hard elastic zones or precipitates within a soft elasto-plastic matrix. The internal stresses originate from the requirement of strain continuity at the interfaces. As a result, each phase is subjected to an average internal stress whose effect is to adjust its total average strain to a common value. Through this dependence on local strains, these internal stresses depend on the applied stress; they do not vanish when a specimen is unloaded after having been plastically deformed. This composite approach to internal stresses has been successfully applied to nickel based superalloys, particularly in the case of [001] stress axes, parallel to the faces of the $\gamma'$ cuboids (Kuhn, Biermann, Ungar and Mughrabi 1991, Mughrabi, Biermann and Ungar 1992).

The present model is much in the spirit of the one by Véron et al. (1996), in the sense that the microstructure of the superalloy is decomposed into a combination of sub-elements which are assumed to deform in series and/or in parallel. Such models are necessarily approximate, as they consider the spatial averages of only one stress component, the one parallel to the stress axis. They also neglect the long-range elastic interactions between the hard precipitates as well as the complex state of stress in the crossings between channels. Although a full elasto-plastic computation of compatibility stresses should be performed for a complete solution to this problem, this was not deemed necessary in the present context. Indeed, for simple orientations of the loading axis, the internal stresses computed within a composite approach are in reasonable agreement with the available data (Véron 1995).

To allow for a comparison with the experimental results described in the introductory part, an unidirectional stress $\sigma_a$ is applied along the [001] axis and only the components of the strain and stress tensors parallel to this axis are estimated. As the applied stress destroys the ternary symmetry of the ideal microstructure, one has to distinguish between two types of $\gamma$ channels, the two sets of "vertical" channels that run parallel to [001] and the "horizontal" channels parallel to the other cube directions. As illustrated by fig. 4, the structure of the superalloy is considered as a periodic array of elementary units consisting of two horizontal sub-units strained in series. One consists of a horizontal $\gamma$ channel (H) and the other is made up of a $\gamma'$ precipitate and a vertical channel (V) strained in parallel.

FIG. 4 HERE

The constitutive forms of the two phases are taken as follows. The "hard" $\gamma'$ phase always behaves in an elastic manner, with a Young's modulus $E$. The two phases are assumed to have same elastic constants. The plastic behaviour of the ductile $\gamma$ phase is characterised by its yield stress $\sigma_o$ and a constant strain hardening coefficient $h$. As far as the lower sub-unit (H in fig. 4-b) is concerned, it is only submitted within the present description to the uniform applied stress. For the mixed sub-unit, each phase must have the same total deformation along the loading axis, in order to satisfy the requirement of strain continuity across the interface. If $\varepsilon_\gamma$ and $\varepsilon_{\gamma'}$ are respectively the total strains parallel to [001] in the $\gamma$ and $\gamma'$ phases of the mixed sub-unit, we have:

$$\varepsilon_\gamma = \varepsilon_{\gamma'} \qquad (2)$$

As long as the magnitude of the applied stress $\sigma_a$ is such that $\sigma_a - \sigma^m < \sigma_{oV}$, where $\sigma_{oV}$ denotes the yield stress of the vertical channels, the two phases deform elastically in a compatible manner. When the $\gamma$ phase enters the plastic domain, compatibility stresses $\sigma_\mu$ and $\sigma_{\mu'}$ set in to counteract its larger deformability. The corresponding local stresses $\sigma_\gamma$ and $\sigma_{\gamma'}$ along a direction parallel to the stress axis, are related to the total strains through the constitutive forms:

$$\varepsilon_\gamma = \frac{\sigma_\gamma}{E} + \frac{(\sigma_\gamma - \sigma_{oV})}{h}; \quad \varepsilon_{\gamma'} = \frac{\sigma_{\gamma'}}{E} \qquad (3)$$

The Albenga's law applies to the average internal stresses i.e., the compatibility stresses. It expresses the fact that the volume average of the internal stresses is zero. In the mixed sub-element, the equivalent volume fractions of the two phases are proportional to *l*, the width of the $\gamma$ channels, and to *a*, the side of the cuboidal precipitates, respectively (cf. fig. 4-b). Thus, we have:

$$l\sigma_\mu + a\sigma_{\mu'} = 0 \qquad (4)$$

In each phase, the total [001] component of the stress tensor is the sum of the applied stress and of the average compatibility stress:

$$\sigma_\gamma = \sigma_a + \sigma_\mu \; ; \; \sigma_{\gamma'} = \sigma_a + \sigma_{\mu'} \qquad (5)$$

Finally, by combining eqs. (2) to (5), one obtains the internal stresses in the two phases:

$$\sigma_\mu = -\frac{1}{1+\frac{h}{E}(1+\frac{l}{a})}(\sigma_a - \sigma_{oV}) \; ; \; \sigma_{\mu'} = -\frac{l}{a}\sigma_\mu \qquad (6)$$

As soon as the vertical channels start deforming plastically, internal stresses $\sigma\gamma$ and $\sigma\gamma'$ develop, which tend to reduce the deformation in the ductile vertical channels and to increase it in the hard precipitates (cf. fig. 5). As shown by eq. 6, the amplitude of these internal stresses increases in proportion to the stress in excess of the yield stress of the $\gamma$ phase. To numerically estimate this composite effect the following values were introduced into eq. (6). The strain hardening coefficient of the $\gamma$ phase, $h = 2.36$ GPa, was taken from values quoted for superalloys similar to the present one and at the same temperature (Jouiad 1996). The Young's modulus can be deduced from the elastic constants at 950 °C (Priester, Khalfallah and Coujou 1997) or from high temperature values quoted in the literature (Courbon 1990, Müller, Glatzel and Feller-Kniepmeier 1992). A reasonable average value is $E = 80$ GPa and $l/a \approx \frac{1}{4}$. Then,

$$\sigma_\mu \approx -0.96(\sigma_a - \sigma_{oV}) \quad \text{and} \quad \sigma_{\mu'} \approx +0.4(\sigma_a - \sigma_{oV}). \tag{7}$$

One sees that these compatibility stresses can become substantial when the vertical channels are strain-hardened.

FIG. 5 HERE

## 4. INTERNAL STRESSES AND SLIP ACTIVATION

The two internal stresses computed so far have quite different origins. However, both of them influence the selection of the active slip systems, the nature of the mobile dislocations as well as their spatial distribution. In this part, these effects and their evolution under strain are discussed in a semi-quantitative manner.

### *4.1 Relaxation of misfit stresses*

The misfit stresses constitute a property inherent to the non-deformed material. However, they can be modified, by the presence of dislocations generated in the $\gamma$ phase and further blocked at the interface between the two phases (Biermann, Kuhn, Ungar, Hammer and Mughrabi, 1991). In the present case, the lattice mismatch induces compressive misfit stresses in the channels (cf. part 2). The non-screw dislocations stored at the interfaces can, then, relax the misfit stresses provided that their extra half-plane are in the $\gamma'$ precipitates. Hence, the misfit stresses, which contribute to the Peach-Koehler force, influence the nature of the mobile dislocations and are also modified by the latter. Previous results in the case of tensile deformation (Carry and Strudel 1977, Fredholm 1987, Buffière and Ignat 1995, Brien and Décamps, submitted) show that the dislocations activated in the channels perpendicular to the stress axis are precisely those that tend to relax the misfit.

A simple reasoning can be used in order to explain this feature. Consider for instance the horizontal channels in [001] tension. Due to the requirement of strain continuity at the γ/γ' interfaces, bi-axial tension stresses are induced in the ductile γ channels. At dislocation scale, these stresses are nothing else than the stress fields of the dislocations in the activated slip systems. Indeed, dislocations are by definition the defects that ensure compatibility between slipped and non-slipped areas in a slip plane (Mura 1993). As the misfit stresses induce a bi-axial compression, the slip systems that are preferentially activated under the effect of compatibility stresses are, in effect, such that they relax the misfit stresses. In order to obtain more quantitative predictions, the active slip systems in the channels are first determined as those submitted to the largest resolved stress, assuming that only the applied stress and the misfit stress are present. As soon as the vertical channels enter the plastic domain, composite stresses are generated which may further modify the nature of the active slip systems. This sequence of activation events is now examined in detail as a function of the type of loading.

### *4.2. Resolved stresses*

In a first step, the local stress is calculated as the sum of the stress applied along the [001] axis, $\sigma_a$, and of the misfit stress tensor $\sigma_m$ as given in part 2. Care is taken to define the resolved stress with its direction and amplitude, in order to analyse the exact nature of the dislocations stored at the interfaces. For each slip system *(s)*, defined by its Burgers vector $b^s$ and its line direction $\ell_u^s$, oriented according to the FSRH convention (Bilby, Bullough and Smith 1955), the Peach-Koehler relation yields the corresponding force $F_{PK}^s$ or stress $\sigma_{PK}^s = \dfrac{F_{PK}^s}{b^s}$ per unit line of dislocation length. The unit normal $n^s$ to the slip plane is defined as $n^s = (\ell^s \times b^s) / |\ell^s \times b^s|$, such that it points from the missing half-plane to the extra half-plane of a dislocation loop (cf. eq. 3-86 of Hirth and Lothe 1992. This property is easily demonstrated by performing a Burgers circuit around an edge dislocation line). The resolved stress $\sigma^s$ is obtained by projecting the Peach-Koehler stress $\sigma_{PK}^s$ deduced from the Peach-Koehler force on the slip plane and along a direction normal to the line (cf. eq. 3-91 of Hirth and Lothe 1992):

$$\sigma^s = \sigma^s_{PK}.(\ell^s_u \times (b^s \times \ell^s_u))/|b^s \times \ell^s_u| = \sigma^s_{PK}.(n^s \times \ell^s_u) \qquad (8)$$

Equation (8) allows to define the direction along which the Peach-Koehler force is resolved. By rewriting the final result in component form, one obtains the resolved stress as measured along this direction (Kosevich, 1979):

$$\sigma^s = (\sum_{i,j} \sigma_{i,j} b^s_j n^s_j) \qquad (9)$$

Denoting by $H$ the horizontal channels and by $V_i$ the vertical ones ($V_1$ is normal to [100] and $V_2$ to [010]), we obtain from eq. 9:

$$\sigma^s_H = b^s_3 n^s_3 (\sigma_a + K|\delta|) \qquad (10)$$

Here, the negative misfit stress is written $-|\delta|$, so that one can better see that its contribution is in the same direction as that of the applied stress. In addition, use was made of the relation $\sum_i b^s_i n^s_i$, that holds between the components of the two orthogonal vectors $b$ and $n$. The slip systems to be activated first upon loading are therefore those with non-zero components of $b$ and $n$ parallel to the stress axis. For the other slip systems, the resolved stress vanishes. In the vertical channels, eq. (9) yields:

$$\sigma^s_{V_i} = \sigma_a b^s_3 n^s_3 - K|\delta|(b^s_3 n^s_3 + b^s_i n^s_i) \qquad (11)$$

Figure 6 shows in the Thompson's tetrahedron representation, the resolved shear stress values obtained for all the octahedral slip systems according to the type of channel they belong to (a projection factor which is common to all stress components is omitted for the sake of simplicity). Figure 7 reproduces these results in real space, illustrating the difference in local stresses in the horizontal and vertical channels due to their different orientation with respect to the applied stress. These results indicate that the number of active slip systems depends on the orientation of the channels with respect to the loading axis, thus leading to the microstructural anisotropy between horizontal and vertical channels recorded at small strains (cf. Fig. 1-a).

FIGS. 6 and 7 HERE

### *4.3 Relaxation of the misfit*

We can now check whether or not the mobile dislocations that are first generated move in such a way as to relax the misfit stresses. Figure 8 shows a dislocation line in a channel, with the slip plane normal oriented according to the convention defined above. The misfit is relaxed only if the resolved stress moves this dislocation towards the left interface, the extra half-plane lying then inside a precipitate. Therefore, misfit stress relaxation occurs if a simple criterion is fulfilled: the normal *n* to the slip plane and the resolved stress on the dislocation line $\tau$ must have components of same sign in the direction of the normal to the considered interface.

FIG. 8 HERE

We consider, as an example, one slip system that is likely to be activated in the horizontal channels according to eq. 10 and fig. 6, the one defined from the line direction $[1\bar{1}0]$ of the mixed segment and the Burgers vector $DB=a/2[011]$. The positive normal to the slip plane is $n = [\bar{1}\bar{1}1]$. From equation (10), with $b_3 > 0$ and $n_3 > 0$ and in tension, the resolved stress is positive when measured along the direction of $n \times \ell = [112]$ (cf. eq. 8). Both the resolved stress and the normal *n* have components of same sign along the [001] direction normal to the horizontal interface. Thus, this slip system effectively relaxes the misfit stress in the horizontal channels in tension. In compression, the direction of the resolved stress is reversed and the opposite result holds.

The same analysis was performed for all the slip systems that are activated in the horizontal channels. As a result, all of them were found to relax the misfit in tension. In the vertical channels, it was found in the same manner that the mixed segments deposited at the vertical [010] or [100] interfaces have either a neutral effect on the misfit or tend to relax it in compression.

A simple estimate shows that 5 ±1 interfacial dislocations are necessary, in the present case, to relax the misfit stress (Brien and Décamps submitted). The local axial plastic strain produced by 5 dislocation loops in a channel is $\varepsilon = nb\ell a^2/a^3 = nb/a$. With $b = 0.358$ nm and $a \sim 400$ nm, the local plastic strain needed to relax the misfit is $\varepsilon \sim 4.5 \cdot 10^{-3}$. With a strain hardening coefficient of about 2.3 GPa (cf. Part 3), this corresponds to a hardening of the gamma phase of $h\varepsilon \sim 10$ MPa. This value is much smaller than the misfit stress $K|\delta| \sim 160$ MPa. Hence the stress increase necessary to relax the misfit stress is small compared to the differences in resolved stresses between the different slip systems in the horizontal channels.

### *4.4 Activated slip systems in tension, compression and low cycle fatigue*

Combining the results obtained in the previous parts, it is possible to predict the nature and number of the active slip systems in each type of channel, vertical or horizontal, and for different types of loading. For this purpose, one has to account for the resolved stresses, for the relaxation of the misfit stresses and for the building up of compatibility stresses. The net result is summed up below in figs. 9 and 10 for the cases of tension and compression, respectively. In each figure, a stress axis indicates the applied stress, whose values range from $\sigma_o^\gamma - K|\delta|$ to $\sigma_o^\gamma + 2K|\delta|$, where $\sigma_o^\gamma$ is the yield stress of the $\gamma$ channels in the absence of internal stress. This yield stress accounts for solute hardening and various confinement effects, as discussed by Nabarro (1995). Several columns are presented which show the number and location of the active slip systems and their evolution with the applied stress. The column RSS refers to the situation at the beginning of yield, i.e., without misfit stress relaxation and compatibility stresses. These results directly derive from fig. 7. The column M shows how the slip pattern is modified by the relaxation of the misfit stress (cf. section 4.3), taking into account that this relaxation occurs within a small stress interval compared to $K|\delta|$. A network of interface dislocations is drawn wherever misfit stress relaxation has occurred. In the third column, M+C, a further modification due to the internal compatibility stresses in the vertical channels is introduced. This contribution is estimated from eq. 7 and resolved on the considered slip systems. Therefore, this column represents the expected experimental behaviour. Finally, the symbols attached to each set of active slip systems allow to qualitatively estimate the slip activity by comparing the effective stress in each slip system to the applied stress. The range of effective stresses can be deduced from the horizontal stress axes in figs. 9 and 10.

FIGS. 9 and 10 HERE

The results obtained in tension can also be used to examine the behaviour in repeated fatigue, i.e., with a maximum applied stress that is cycled between zero and a maximum tensile stress $\sigma_a$. The relaxation of misfit stresses in tension or repeated fatigue is conditioned by the occurrence of a small plastic strain ($\varepsilon \sim 4.5\ 10^{-3}$) in the horizontal channels. This strain corresponds to a macroscopic cumulated strain and a macroscopic flow stress value for the superalloy that is slightly above the stress at which the horizontal channels yield. The tension results correspond then, to the limit case of repeated fatigue with only one cycle ($N = 1$). The predicted behaviour in repeated fatigue with $N > 1$, is shown in the right column of fig. 9. Interfacial dislocations are expected to appear under smaller maximum stresses, after a certain critical number of cycles $N_o$ that increases with decreasing maximum stress. This is precisely what is experimentally observed and $N_o$ is found to be in the range 1-25 in the low cycle fatigue of the AM1 superalloy at high temperature (Brien and Décamps submitted). Of course, the minimum stress for which misfit relaxation occurs necessarily corresponds to the critical stress for slip activation in the horizontal channels. In all these cases, the horizontal channels are the first ones that yield plastically. The corresponding applied stress is $\sigma_o^\gamma - K|\delta|$, which is then the yield stress of the superalloy. When slip is activated in the vertical channels, for $\sigma_a = \sigma_o^\gamma$, the

latter are always less stressed in comparison, so that plastic deformation preferentially occurs in the horizontal channels. Under larger applied stresses, typically $\sigma_a > \sigma_o^\gamma - K|\delta|$ or cumulated plastic strains, the deformation tends to become more uniform (cf. Fig. 1).

The activated slip systems are such that they relax the internal stresses in the (001) horizontal interfaces but not in the two vertical ones, (100) and (010). As a consequence, the elastic gradients induce type N coarsening. Within the present model, the same conclusion should hold under high applied stresses, since this should not modify the sign of the elastic gradients. Experimental observations on the microstructures developed during repeated fatigue show, however, that type N coarsening does not occur at high stresses (cf. fig. 1-b). In such conditions and in agreement with the observations mentioned above, one has to conclude that cross-slip and climb are activated. Indeed, these stress-assisted mechanisms can relax the internal stresses at the vertical interfaces, thereby inducing the vanishing of the corresponding elastic gradients. Therefore the experimental observation of non-directional coarsening under high stress in repeated fatigue can be explained only via the activation of additional relaxation mechanisms, viz. cross slip and climb of dislocations (Brien and Décamps, submitted).

In compression (cf. fig. 10), the present model also predicts an uniformisation of the slip activity with increasing applied stress. A comparison of the results depicted in figs. 9 and 10 shows that the microstructures are expected to be more anisotropic in tension than in compression, for a given absolute value of the applied stress.

FIG. 11 HERE

The case of alternate fatigue is more complex. If the mobile dislocations tend to relax the misfit stresses at some interfaces during one half cycle, the effect is reversed during the next half cycle (cf. 4.3) and the dislocations which reach the same interfaces tend to increase the misfit stress. However, due to the differences in slip activity in tension and compression, the two effects do not always compensate each other exactly. The situation is summed up in fig. 11 which reproduces the predicted slip patterns during the first half cycle in tension or compression and the net result that is predicted in alternate fatigue after one fully reversed cycle. This is sufficient to obtain a good indication of the expected behaviour since precipitate coarsening is found to occur during the very first cycles (Brien, 1995). The cases where the stress reversal induces exactly opposite slip activity and no global effect on misfit stress relaxation are indicated by the symbol *n* (for neutral). The arrows indicate whether these dislocations relax or reinforce the misfit stresses. The relaxation of the misfit stress in the vertical channels is only one-dimensional (1D), whereas it is two-dimensional (2D) in horizontal ones. As a consequence, there are elastic energy gradients in alternate fatigue, whose sign does not depend on the maximum value of the applied stress and is the same as in repeated fatigue. This explains the occurrence of type N coarsening in alternate fatigue.

## 5. CONCLUSION

In the present work a simplified model is proposed that allows to estimate the two main sources of internal stresses present in a superalloy with high volume fraction of ordered precipitates, the misfit stresses and the compatibility stresses. As a consequence, the nature of the slip systems activated at yield, the resulting spatial anisotropy of slip activity and the relaxation of misfit stresses have been investigated. Under increasing applied stress, the sequence of activation of slip systems in the horizontal and vertical channels in tension, compression and fatigue can be predicted in good agreement with the available experimental results, as well as the formation of interfacial dislocation networks. The observed evolution from anisotropic to more uniform dislocation microstructures in tension and compression also comes out from the complex interplay between the various sources of local stress. In addition, the present model explains why the two types of low cycle fatigue, repeated and alternate, lead to heterogeneous microstructures of deformation in superalloys and to type N coarsening. More generally, the present model offers the possibility of estimating in simple terms the gradients of elastic energy that contribute to anisotropic precipitate coarsening. This work also shows that the model by Véron et al. (1996) for coarsening in monotonic deformation only applies to the case of repeated fatigue at low stresses. In the other cases (large stresses and alternate fatigue), a more detailed examination of the elastic gradients is necessary.

## ACKNOWLEDGEMENTS

VB and BD wish to thank SNECMA and CNRS for their financial support through the CPR "Stabilités Structurales des Superalliages Monocristallins". LK acknowledges the support of the National Science Foundation under grant No PHY 94-07914. This work was carried out at LMS, URA CNRS 1107, Université Paris-Sud, Bâtiment 413, 91405 Orsay Cedex and at LEM, Unité Mixte CNRS-ONERA, UMR 104 CNRS, BP 72, 92322 Châtillon Cedex in France.

**FIGURES**

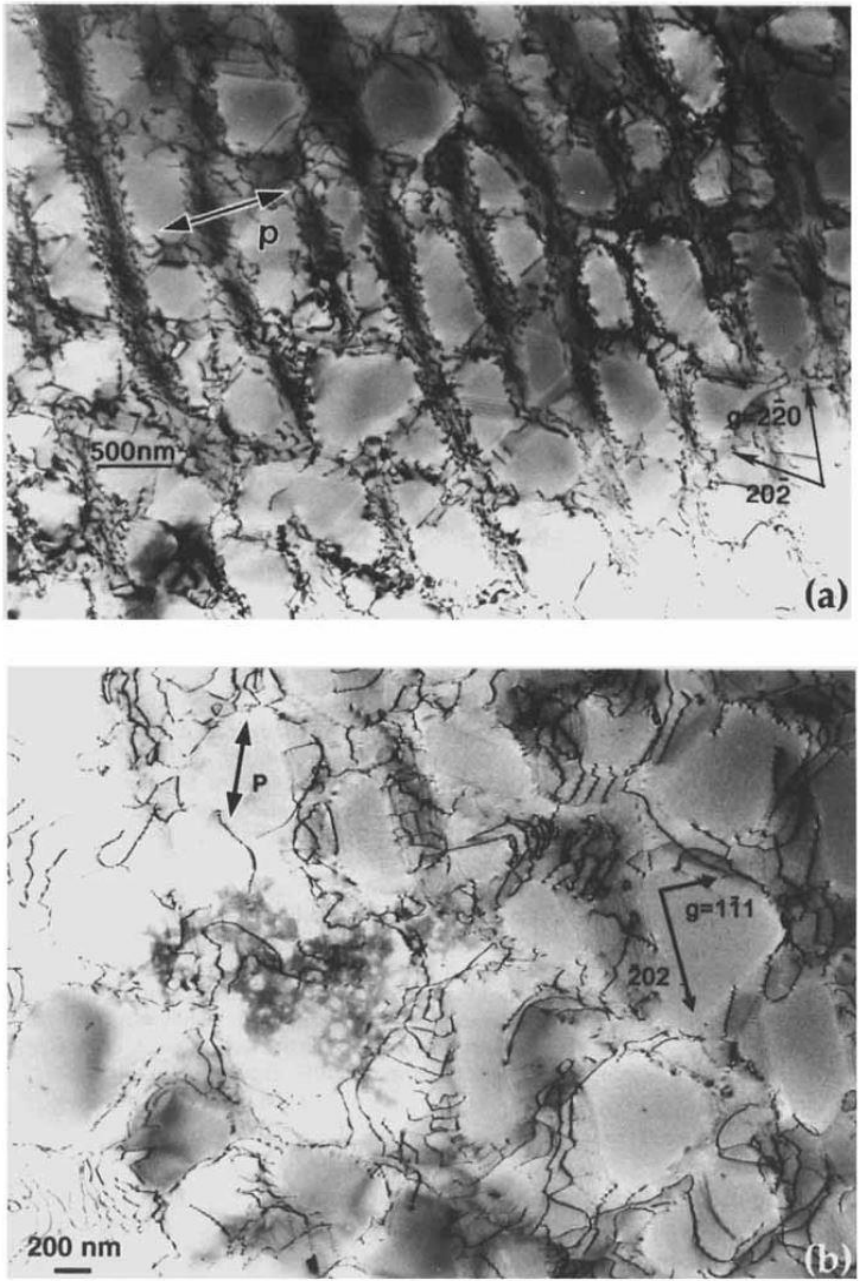

Figure 1. Transmission electron micrograph showing the dislocation microstructure of the *AM1* superalloy cycled in repeated tensile fatigue (T = 950 °C, [001] stress axis) with a total strain amplitude of 1.3 % per cycle. Influence of the cumulated strain. a) Low stress, 25 cycles ([111] zone axis). The channel perpendicular to [001] (arrow) contains a significantly larger density of dislocations than the two other channels. This microstructure leads to type N coarsening of the γ' precipitates. b) High stress, 1300 cycles. Equivalent densities of dislocations are found in the three types of channels ([121] zone axis). This microstructure leads to isotropic coarsening of the γ' precipitates.

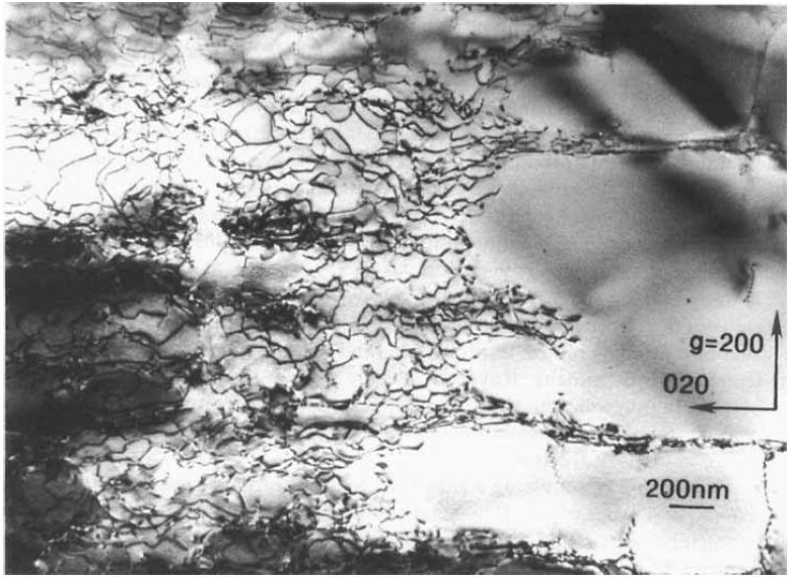

Figure 2. Dislocation microstructure of the *AM1* superalloy in alternate fatigue (T = 950 °C, [001] stress axis) after 113 cycles with a total strain amplitude of 1.3 % per half-cycle. The presence of dislocation nets along the [001] channels, that are here parallel to the foil plane, is associated with type N coarsening.

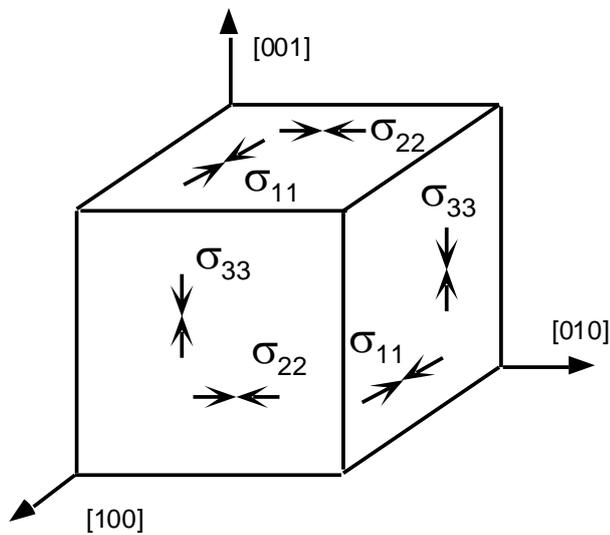

Figure 3. Schematic representation of the misfit stresses in the $\gamma$ channels for a negative value of the lattice mismatch.

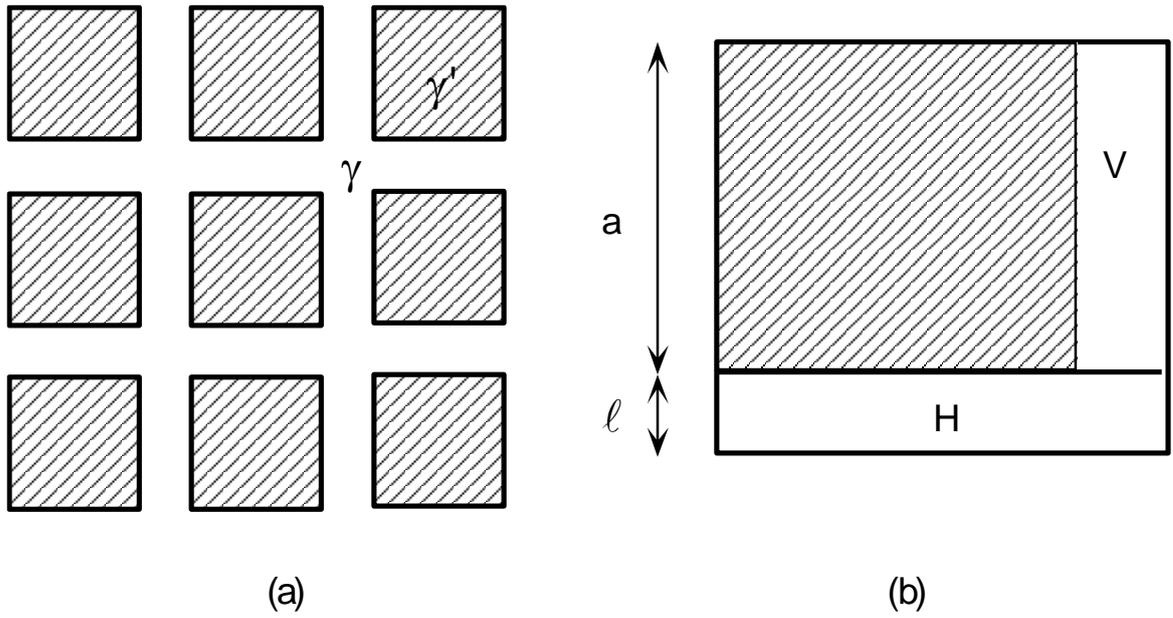

Figure 4. Composite model of the biphased γ/γ' superalloy. (a) - Ideal microstructure, (b) - elementary cell of the composite model.

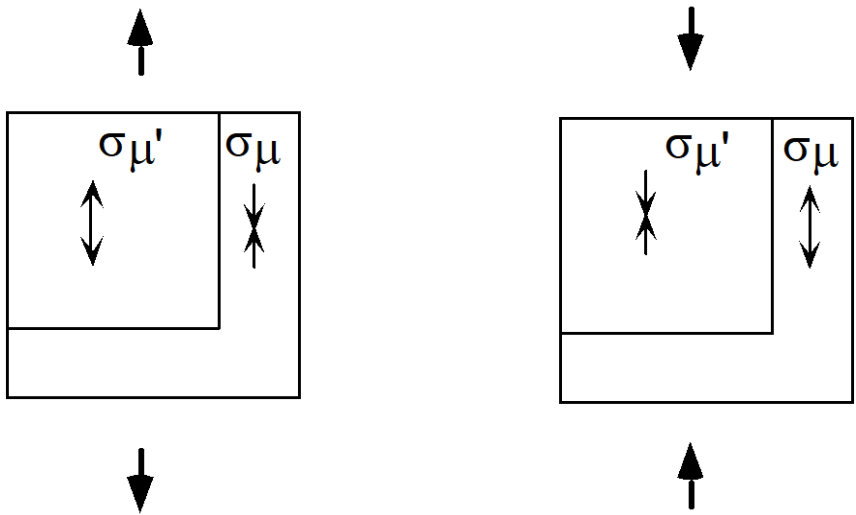

Figure 5. Compatibility stresses in the composite model for the biphased superalloy in [001] tension or compression.

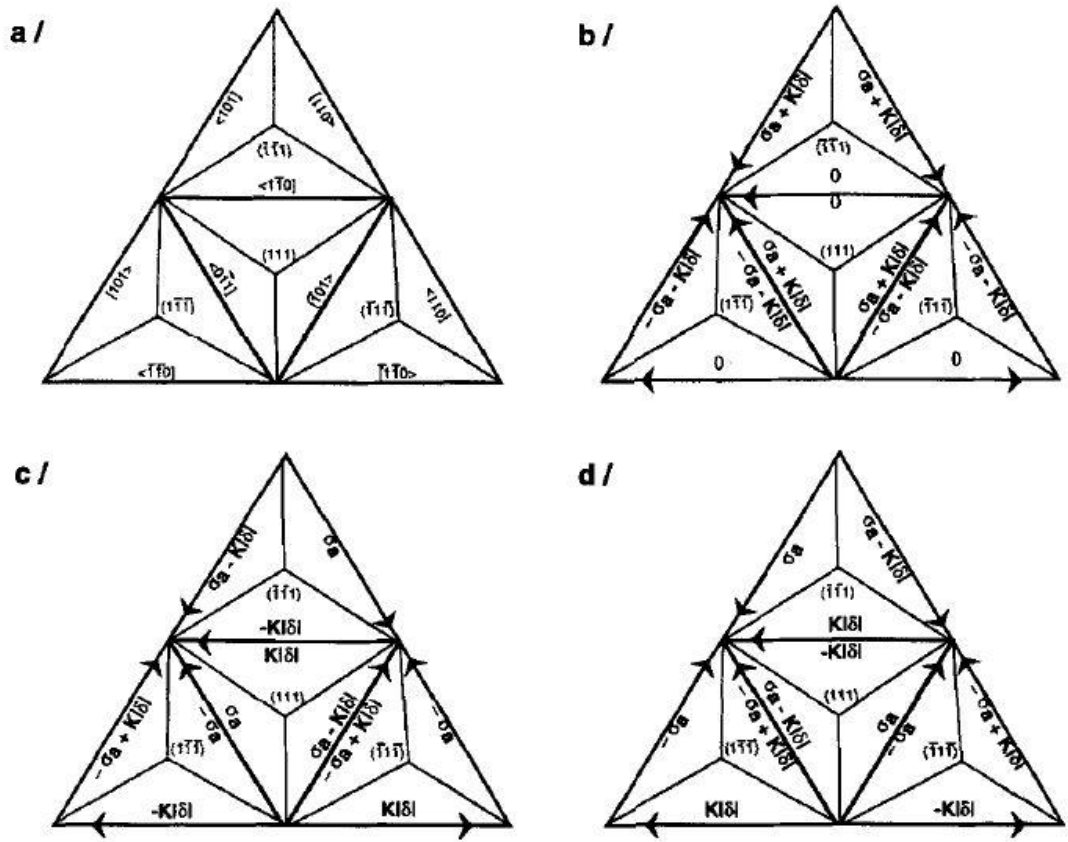

Figure 6. Resolved shear stresses in the presence of misfit stresses for all the octahedral slip systems in the γ channels (all stresses are multiplied by a Schmid factor $F = 1/\sqrt{6}$). (a) - Slip systems in the Thompson's tetrahedron. The resolved shear stresses are indicated for (b) - the [001] horizontal channel, (c) - the [100] vertical channel and (d) - the [010] vertical channel.

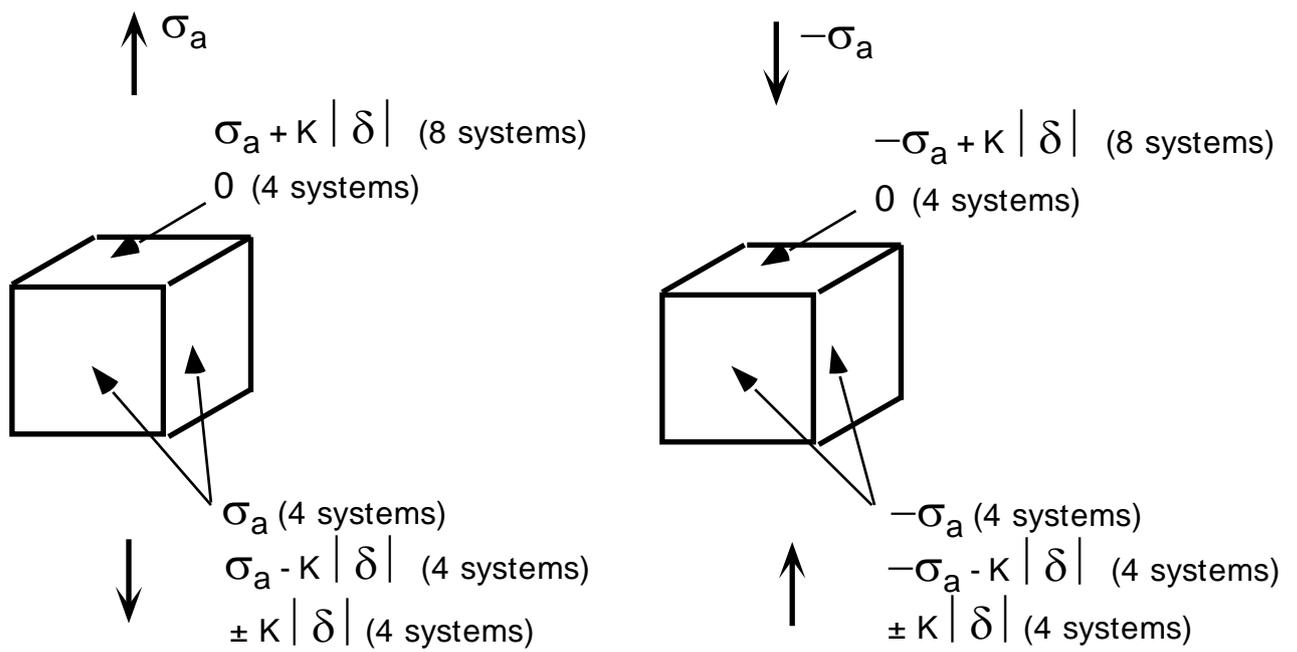

Figure 7. Resolved applied plus misfit stresses on the octahedral slip systems of the $\gamma$ phase in tension and compression (all stresses are multiplied by a Schmid factor $F = 1/\sqrt{6}$).

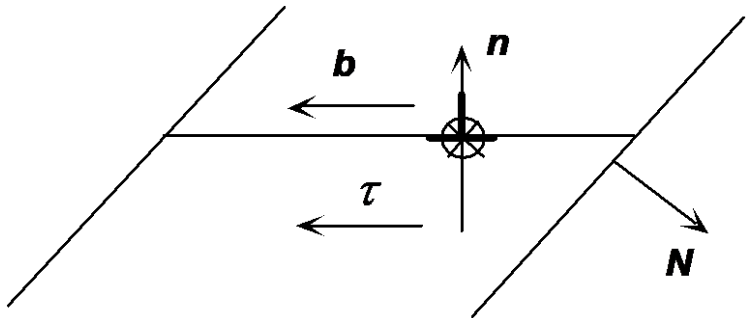

Figure 8. Schematic view of a non-screw dislocation in a $\gamma$ channel of normal $N$. The extra half-plane is such that it relaxes the misfit if the resolved stress moves the dislocation towards the left.

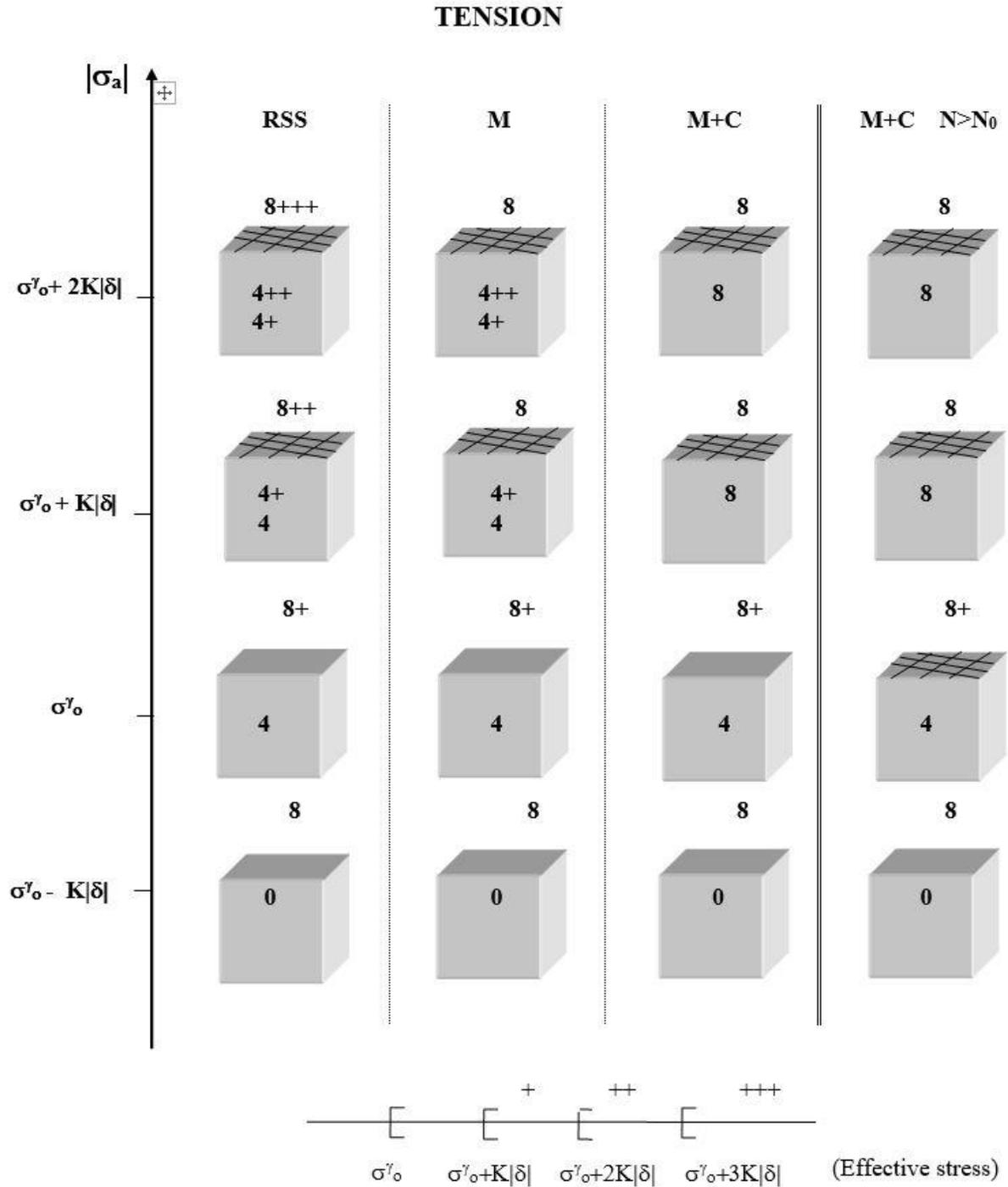

Figure 9. Active slips systems in the horizontal and vertical channels in [001] tension as a function of the applied stress $\sigma_a$ (indicated in the vertical axis at left). The column RSS shows the number and location of the active systems under the effect of the applied stress and the non-relaxed misfit stresses. The column M shows how this result is modified by the relaxation of misfit stresses (the relaxed interfaces are materialised by a network of dislocations) and the column M+C how it is further modified by compatibility stresses in the vertical channels. The symbols (+) indicate the effective (non-resolved) stresses in the channels, according to the convention shown in the horizontal axis. The column M+C ($N>N_o$) refers to repeated fatigue (see text for detail).

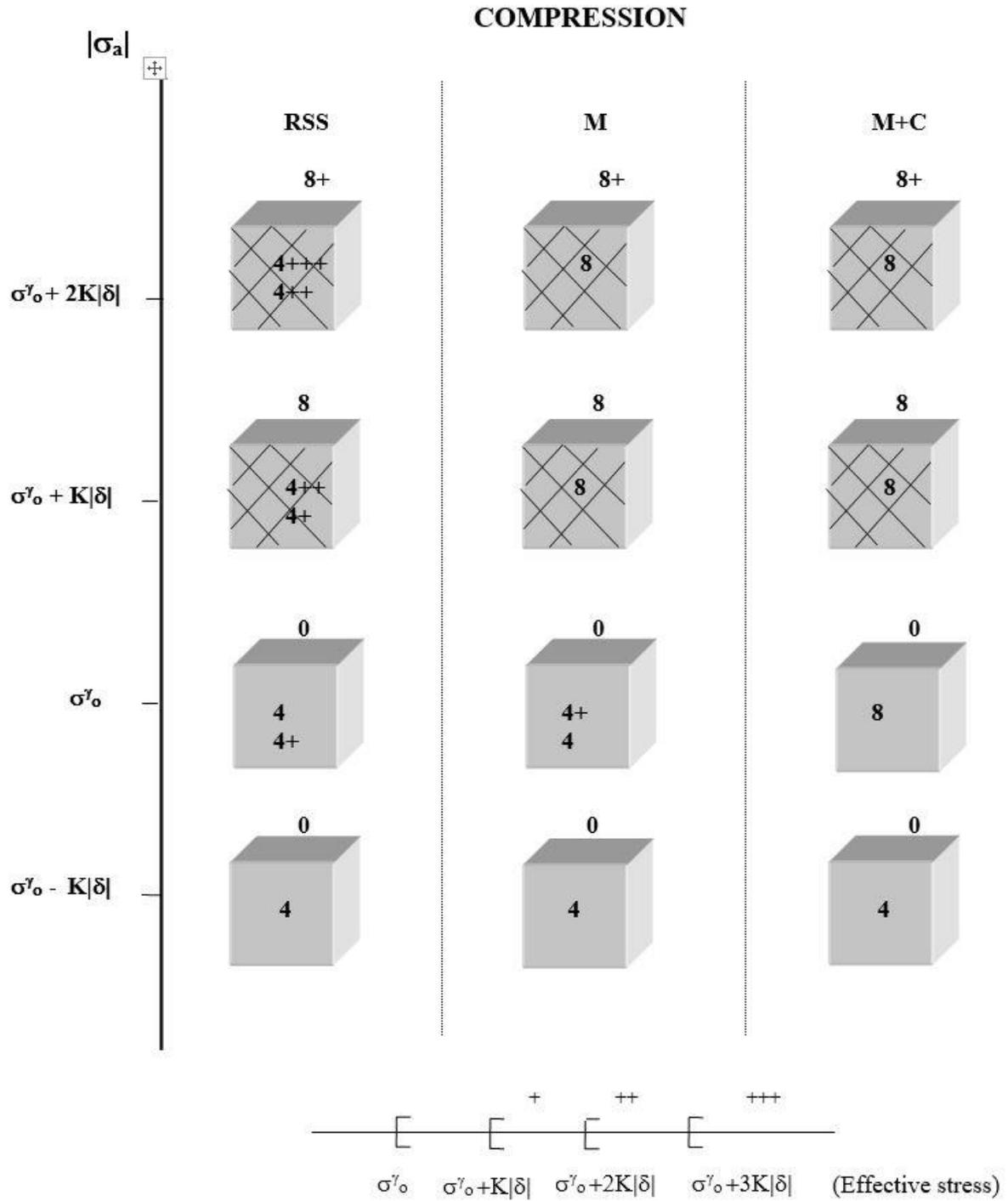

Figure 10. Same as figure 9, [001] compression.

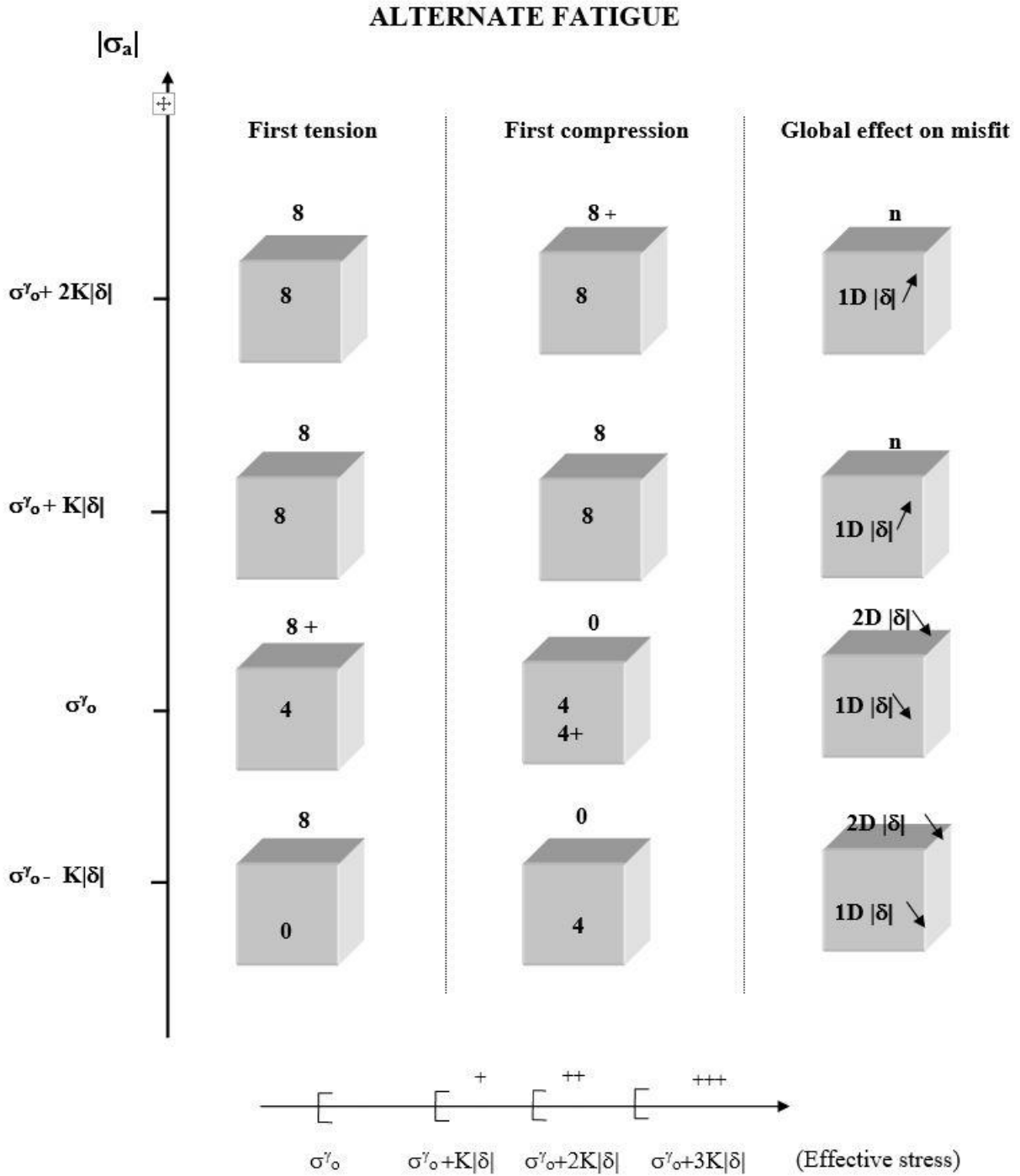

Figure 11. Same as figures 9 and 10, alternate fatigue. The first two columns at left show the predicted slip activity at the end of the first half-cycle in tension or compression, respectively and the column at right shows the net result after one cycle, with a first half-cycle in tension. The symbols 1D and 2D indicate the number of distinct types of interfacial dislocations and the arrow indicates whether these dislocations relax or enhance the misfit stresses.